\documentstyle[amssymb,11pt]{article}
\def\ltsim{\lower3pt\hbox{$\, \buildrel < \over \sim \, $}}
\def\gtsim{\lower3pt\hbox{$\, \buildrel > \over \sim \, $}}
\def\indic{\lower3pt\hbox{$\, \buildrel (m_1,m_2)\not=0 \over
\kappa_{\vec m}\ll 1 \, $}}
\begin{document}
\begin{flushright} 
{OUTP-99-34-P}\\  
\end{flushright} 
\vskip 0.5 cm
\begin{center}
{\Large {\bf String thresholds and Renormalisation 
Group Evolution \bigskip }}\\[0pt]
\bigskip {\large {\bf Dumitru Ghilencea\footnote{{{{E-mail:
D.Ghilencea1@physics.oxford.ac.uk}}}} } and {\bf Graham 
Ross\footnote{{{{E-mail: G.Ross1@physics.oxford.ac.uk}}}}\bigskip
}}\\[0pt]
{\it Department of Physics, Theoretical Physics, University of Oxford}\\[0pt]
{\it 1 Keble Road, Oxford OX1 3NP, United Kingdom}\\[0pt]
\bigskip

\vspace{2cm} $Abstract$
\end{center}

{\small We consider the calculation of threshold effects due to Kaluza Klein
and winding modes in string theory. We show that for a large radius of
compactification these effects may be approximated by an effective field
theory applicable below the string cut-off scale. 
Using this formalism we show that
the radiative contribution to gauge couplings involving only massive Kaluza
Klein and winding modes may be calculated to all orders in perturbation
theory and determine the full two loop contribution involving light modes
and estimate the magnitude of the higher-order contributions. 
For the case of the weakly coupled heterotic string we also
discuss how an improved calculation can be made incorporating the 
string theory threshold corrections which avoids the limitations of
the  effective field theory approach. Using this formalism we
determine the implications for gauge coupling unification for one
representative model including the effects of two loop corrections above the
compactification scale. Finally we discuss the prospects for gauge
unification in Type~I models with a low string scale and point out potential
fine tuning problems in this case. }

\newpage

\section{Introduction}

The possibility of large spatial dimensions \cite{antoniadis1} in addition
to the four dimensional Minkowski space-time has recently received
considerable attention, particularly within the context of the unification
of the gauge couplings in some string models 
\cite{dienes_v1,extradim,mondragon,mohapatra,carone,bogdan,zurab3}. In such
models, the presence of power-law \cite{dienes_v1,veneziano,marco}
``running'' of the gauge couplings induces a significantly different picture
from the familiar logarithmic running normally found in four dimensional
field theory. From the four dimensional point of view the different running
of the gauge couplings is due to the presence in the spectrum of the
additional excitations (Kaluza-Klein states) associated with the extra
spatial (compact) dimensions. Alternatively it may be viewed from the higher
dimensional point of view as simply due to the different behaviour of
propagators in higher dimensions.

Due to the point-like nature of the couplings in field theory, the radiative
corrections due to the Kaluza Klein states are intrinsically ill defined
because the infinite sum over the Kaluza Klein tower diverges. To make sense
of these corrections it is necessary to have a regularization procedure such
as is provided by the string. For the closed string the two dimensional
structure of the string worldsheet means that, at distance scales at which
string excitations are relevant, the couplings are no longer point-like and
indeed have a formfactor which falls exponentially fast at short distances
above the string scale. For the open string the situation is somewhat
different because states far above the string scale can contribute 
\cite{dudash, bachas,antoniadis_bachas,fabre}. 
However, due to supersymmetry only
massless $({\cal N}=1)$ states (with cut-off at the string scale) and
massive $({\cal N}=2)$ states (with cut-off at the largest compactification
scale) contribute. As a result, in both cases only a finite number of Kaluza
Klein or winding states (KKW) make significant contributions to the
effective low energy theory. For this reason it is possible to determine the
effect of the KKW states in an effective field theory approach, applicable
at scales below the cut-off scale. In this approach the radiative
corrections due to those KKW states with mass less than the cut-off scale
are calculated in the effective field theory, while the radiative
corrections of the remaining KKW states are included in the boundary
condition for the operators of the effective theory. This formalism provides
a convenient method for determining perturbative corrections to gauge and
other couplings.

While the string gives finite, and in principle calculable, radiative
corrections due to the massive states of the theory, it does not guarantee
that the perturbative calculation converges even if the couplings, $\alpha
_{i}$, of the effective field theory are small. The reason is that, if the
scale of compactification is far below the string scale, there is a very
large number, $N$, of light KKW states. Given this the naive condition for
convergence of the perturbative series is that $\alpha _{i}N$ be small.
However, as we will discuss, due to supersymmetry, the situation is somewhat
better. We show that the one loop contribution of massive KKW modes is
perturbatively exact and calculate it in the limit of large compactification
scale. We also discuss the perturbative contribution involving massless
modes and determine the effect on this contribution of the massive modes at
two loop order. This contribution is not perturbatively exact and gives an
inherent uncertainty to the calculation. We calculate the correction to the
inverse couplings at two loop order and show that the leading contribution
occurs at $\ O(\alpha _{i})$ relative to the leading order which is ${\cal O}%
(N)$. Higher order contributions occur at order ($\alpha _{i}N)^{m}$
relative to this two loop contribution and we discuss the implications of
these terms for the calculability of the couplings.

The outline of this paper is as follows. We first discuss the radiative
corrections in string theories and develop the effective field theory
approach to the calculation of corrections to the gauge couplings of the
theory. In Section \ref{string} we discuss the string regularization of the
contribution of towers of massive Kaluza Klein and winding states in the
context of the weakly coupled heterotic string and also in the context of
Type I (Type I$^{\prime }$) string theories which may admit a low string
scale. We discuss how the contribution of these states may be calculated in
the context of an effective field theory applicable below the string
cut-off  scale
and determine the limits of its applicability. In Section \ref{eft} we
calculate the radiative corrections to gauge couplings using the effective
field theory approach and discuss the convergence of the series in the
presence of the KKW states. In Section \ref{hs} we determine the two loop
radiative corrections to gauge couplings using the full string threshold
calculation, for the case of the weakly coupled heterotic string. The
advantage of the full string calculation is that the boundary conditions for
the effective field theory are determined and we use the heterotic string
example to illustrate the relevance of these terms for the determination of
gauge couplings. In Section \ref{models} we apply the methods developed here
to the analysis of one specific model. Finally in Section 6 we discuss
potential fine tuning problems in (Type I) models with a low string scale.

\section{String threshold corrections\label{string}}

As we have noted the inclusion of the effects of infinite towers of Kaluza
Klein and winding modes associated with new compact dimensions is only well
defined in the context of string theories. Two classes of string threshold
corrections have been explored. The first is for the weakly coupled
heterotic string theory. For it threshold corrections due to Kaluza-Klein
and winding states have been under extensive study both for orbifolds and
smooth manifold compactifications \cite{choi,kap,dixon,kapl,het}. In this
class of string theory the relation of the string scale to the Planck scale
is given by \cite{kap} 
\begin{equation}
M_{s}=\frac{2\,e^{(1-\gamma _{E})/2}\,3^{-3/4}}{\sqrt{2\pi \alpha ^{\prime }}%
} \cong 0.527\,g_{s}\times 10^{18}\,GeV  \label{mstring}
\end{equation}
where $g_{s}$ is the string coupling at the unification. Given the
discrepancy of this scale with the gauge unification scale, $M_{G}\approx
3\times 10^{16}GeV,$ found by continuing the gauge couplings up in energy
using the Minimal Supersymmetric Standard Model (MSSM) spectrum, it is
clearly of importance to determine the threshold effects to see if they may
explain this discrepancy \cite{nillesetal}.

The second class is the Type I (Type I$^{\prime }$) string at weak coupling.
In this case eq.(\ref{mstring}) no longer applies and we have instead the
relation \cite{witten} 
\begin{equation}
M_{s}\sim g\,M_{P}\,e^{-\phi }  \label{witten}
\end{equation}
where $g$ is the gauge coupling, $\phi $ is the dilaton and $M_{P}$ is the
Planck mass. Such a relation allows for a low string scale through the
choice of the dilaton v.e.v. $<\phi >$ and this has caused much interest for
it may bring \cite{witten} the string scale prediction into agreement with
the gauge unification scale, $M_{G}.$ It was further noticed \cite{lykken}
that the mechanism can actually be applied to lower the string scale even
further, perhaps even down to the ``TeV region''. In such scenarios one may
therefore have a low compactification scale \cite{dienes_v1,savas} and a low
string scale as well. At first sight, to preserve the unification of the
gauge couplings, the presence of a such a low string scale requires a
significant change in the running of the gauge couplings, this change being
accounted for by the presence of additional thresholds associated with the
extra spatial dimensions. The threshold corrections to the gauge couplings
have been calculated for various Type I string models in
\cite{dudash,bachas} to give either  a (linear) power-law running or a 
logarithmic running. As we shall discuss in Section \ref{logrun} 
the second case raises the  possibility for the gauge
couplings to unify at a much larger scale than the string \cite{bachas}, 
perhaps even at the original unification scale ($\approx 10^{16}$ GeV).

To discuss the threshold effects it is necessary to review briefly some of
the details of the string calculation. In string theory the form of the
gauge couplings is 
\begin{equation}
\alpha _{a}^{-1}(Q)=k_{a}\alpha _{string}^{-1}+\frac{b_{a}}{2\pi } \log 
\frac{M_{s}}{Q}+\Delta _{a}  \label{gauge}
\end{equation}
where $k_{a}$ is the Kac Moody level, $\Delta _{a}$ is the string
threshold correction (a function of the moduli fields) and $b_a$ is 
the one-loop contribution of the light modes (${\cal N}=1$ sector).

\subsection{Weakly coupled heterotic strings}\label{wchstrings}

We consider first the case of the weakly coupled heterotic string in which
six of the dimensions are compactified on an orbifold, $T^{6}/G$. In such
models the spectrum splits into ${\cal N}=1,$ ${\cal N}=2$ and ${\cal N}=4$
sectors, the latter two associated with a $T^{2}\times T^{4}$ split of the $%
T^{6}$ torus. Due to the supersymmetric non-renormalisation theorem, the $%
{\cal N}=4$ sector does not contribute to the running of the couplings. The $%
{\cal N}=1$ sector gives the usual running associated with light states but
does not contain any moduli dependence. The latter comes entirely from the $%
{\cal N}=2$ sector. For the heterotic string all states are closed string
states and at one loop the string world sheet has the topology of the torus $%
T^2$. For the case of a six-dimensional supersymmetric string vacuum
compactified on a two torus $T^{2}$ the string correction takes the form 
\cite{kap,dixon}. 
\begin{equation}
\Delta _{i}=\frac{\overline{b}_{i}}{4\pi }\int\limits_{\Gamma }\frac{d\tau
_{1}d\tau _{2}}{\tau _{2}}\left( Z_{torus}-1\right)  \label{gag}
\end{equation}
with 
\begin{equation}
Z_{torus}=\sum_{n_{1,2},m_{1,2}\in Z}exp\left[ 2\pi i\tau
(m_{1}n_{1}+m_{2}n_{2})\right] exp\left[ -\frac{\pi \tau _{2}}{T_{2}U_{2}}
|TUn_{2}+Tn_{1}-Um_{1}+m_{2}|^{2}\right]  \label{stringthreshold}
\end{equation}
Here $T\varpropto R_{1}R_{2}$ and $U\varpropto R_{1}/R_{2}$ where $T$, $U$
are moduli and $R_{1}$, $R_{2}$ are the radii associated with $T^{2}.$ For
the case of a two torus $T^{2}$ of common internal radius $T=iT_{2}$ (the
subscript 2 denotes the imaginary part) and $U=i$. Making the dimensions
explicit $T_{2}$ should be replaced by $T_{2}\rightarrow T_{2}^{o}/(2\alpha
^{\prime })\equiv 2R^{2}/(2\alpha ^{\prime })$ with $\alpha ^{\prime }$ as
in eq.(\ref{mstring}). The sum $m_{1},$ $m_{2}$ is over the Kaluza Klein
modes associated with $T^{2}\,$ and the sum $n_{1},$ $n_{2}$ is over the
winding modes. In eq.(\ref{gag}) $\Gamma =\{\tau _{2}>0,|\tau
_{1}|<1/2,|\tau |>1\}$ is the fundamental domain and $\tau =\tau _{1}+i\tau
_{2}$ is the modulus of the world sheet torus. Performing the sums and doing
the integral over the torus world sheet gives the string corrections
$\Delta _{i}^{string}$ of the following form \cite{dixon} 
\begin{eqnarray}
\Delta _{i}^{string} &=&-\frac{\overline{b}_{i}}{4\pi }\ln \left\{ \frac{%
8\pi e^{1-\gamma _{E}}}{3\sqrt{3}}\,{|}\eta (i){|}^{4}\, \frac{T_{2}^{o}}{%
2\alpha ^{\prime }}\left| \eta \left( \frac{iT_{2}^{o}}{2\alpha ^{\prime }}
\right) \right| ^{4}\right\} \label{t2form}\\
&=&-\frac{\overline{b}_{i}}{4\pi }\ln \left\{ 4\pi ^{2}\,{|}\eta (i){|}
^{4}\left( \frac{M_{s}}{\mu _{0}}\right) ^{2}\left| \eta \left[ \frac{i\,3 
\sqrt{3}\pi }{2e^{1-\gamma }}\,\left( \frac{M_{s}}{\mu _{0}}\right) ^{2} %
\right] \right| ^{4}\right\}  \label{di}
\end{eqnarray}
where $\overline{b}_{i}$ is the beta coefficient 
associated with the ${\cal N}=2$ multiplets, 
$M_{s}$ is the string scale, $\mu _{o}\equiv 1/R$, $\eta
(x) $ is the Dedekind eta function and we have replaced $\alpha ^{\prime }$
in terms of $M_s$ (eq.(\ref{mstring})) \cite{kap}. For 
large\footnote{Note that $T_2\approx 5.5 (M_s R)^2$ in ${\overline {DR}}$
scheme, so one can easily have $T_2\approx 50-100$ while $R$ is still
close to the string length scale, to preserve the weakly coupled
regime of the heterotic string. In this section ``large'' $R$
corresponds to values of  $T_2$ in the above range. }
$T_2$ the eta function is dominated by the leading exponential 
\footnote{The Dedekind function is defined by $\eta (T)=e^{\pi
iT/12}\prod_{k=1}^{\infty }\left( 1-e^{2\pi ikT}\right) $} and so one finds
the power law behaviour $\Delta _{i}\propto T_{2}\propto R^{2}$ which has a
straightforward interpretation as being due to the decompactification
associated with $T^{2}.$ Note that, due to the fact that the ${\cal N}=4$
states associated with the $T^{4}$ compactification of four of the six
dimensions do not contribute to the running of the couplings, the power law
behaviour only corresponds to the decompactification of two of the six
compact dimensions.

It is of interest to determine which contributions in eq.(\ref
{stringthreshold} ) dominate in this limit. Using eq.(\ref{stringthreshold})
in eq.(\ref{gag}) one sees the contribution of states more massive than the
string scale are exponentially suppressed. This is in contrast to the
(regulated) field theory result in which all massive states contribute with
a logarithmic dependence on the mass. The difference is due to the absence
of point-like coupling in the string theory corresponding to the two
dimensional distribution of eq.(\ref{gag}). (The field theory result
corresponds to taking the lower limit for the $\tau _{2}$ integration to be $%
0$ rather than $\sqrt{(1-\tau _{1}^{2})}$ of the string). As a result, at
large $T_2$/radius\footnote{As we will see later $T_2\approx {\cal
O}(1)$  is already large enough for our purposes, see also footnote 3.}, 
only the Kaluza Klein modes with mass less than the string
scale make significant contributions. It is this fact that allows us to
reformulate the calculation of the threshold corrections in terms of an
effective field theory calculation valid below the string scale. To make
this point more explicitly we note that the momentum modes alone give the
contribution 
\begin{equation}
\Delta _{i} =\frac{\overline{b}_{i}}{4\pi }\int\limits_{-1/2}^{1/2}d\tau
_{1}\int\limits_{\sqrt{(1-\tau _{1}^{2})}}^{\infty }d\tau _{2}\frac{1}{\tau
_{2}}\left\{ \sum_{m_{1},m_{2}\in Z}exp\left[ -\frac{\pi \tau _{2}}{T_{2}}%
\left( m_{1}^{2}+m_{2}^{2}\right) \right] -1\right\} \label{mom} 
\end{equation}
It is straightforward to compare the result of directly evaluating eq.(\ref
{mom}) with the {\it full} string result of eq.(\ref{t2form}). For example
at $T_{2}=2$ the Kaluza Klein states alone give $99.5\%$ of the {\it full}
threshold correction\footnote{To see this replace the series under the 
integral (\ref{mom}) by their sum, elliptic theta function, 
$\vartheta^2_3(0,e^{-\pi \tau_2/T_2})$.} (\ref{t2form}). 
This means that one may determine the threshold
corrections to an excellent accuracy simply by including the contribution of
states at or below the string scale. To determine how well this contribution
is approximated by the field theory result we  restrict 
the sums over $m_{1},m_{2}$ to a finite number of terms
and  find after some algebra\footnote{We use the result 
\[
E_{1}(z)\equiv \int_{z}^{\infty }\frac{dt}{t}e^{-t}=-\gamma _{E}-\ln
z-\int_{0}^{z}\frac{dt}{t}(e^{-t}-1)\approx -\gamma _{E}-\ln
z\,\,\,if\,\,z\ll 1
\]}
\begin{eqnarray}
\Delta _{i}&=&
\frac{\overline{b}_{i}}{4\pi }\sum_{(m_{1},m_{2})\not=(0,0)}
\int\limits_{-1/2}^{1/2}d\tau _{1}\,E_{1}
\left[ \kappa _{\vec{m}}\sqrt{1-\tau
_{1}^{2}}\right]   \label{momentum}\\
&=&
\frac{{\overline{b}_{i}}}{2\pi }\sum_{(m_{1},m_{2})
\not=(0,0)}^{{}}\ln \frac{M_{s}}{\mu _{0}|{\vec{m}}|}
-\frac{{\overline{b}_{i}}}{4\pi }\sum_{(m_{1},m_{2})
\not=(0,0)}^{{}}\int\limits_{-1/2}^{1/2}d\tau
_{1}\int\limits_{0}^{\kappa _{\vec{m}}\sqrt{1-\tau _{1}^{2}}}
\frac{dt}{t}(e^{-t}-1)  \label{fullsum}
\end{eqnarray}
where 
\begin{equation}
\kappa _{\vec{m}}\equiv \frac{\pi |{\vec{m}}|^{2}}{T_{2}},\,\,\, 
|{\vec{m}}|^{2}=m_{1}^{2}+m_{2}^{2}
\label{kapa}
\end{equation} 
We denote by $\tilde{\Delta}_{i}$ the second term 
in (\ref{fullsum})
 and we will 
restrict both sums in (\ref{fullsum}) to terms with
\begin{equation}
\kappa _{\vec{m}}\equiv \frac{\pi |{\vec{m}}|^{2}}{T_{2}}\ll 1
\label{kapall1}
\end{equation}
First we have that
\begin{equation}
\Delta _{i}(\kappa _{\vec{m}}\leq 1)=\frac{{\overline{b}_{i}}}{2\pi }%
\sum_{(m_{1},m_{2})\not=(0,0)}^{\kappa _{\vec{m}}\leq 1}\ln \frac{M_{s}}{\mu
_{0}|{\vec{m}}|}-{\tilde{\Delta}}_{i}(\kappa _{\vec{m}}\leq 1)
\label{stringcutoff}
\end{equation}
The term ${\tilde{\Delta}_{i}}$ is negligible in the limit (\ref{kapall1})
and we are therefore left with the logarithmic terms. These are just those
of the effective field theory approach, as we will see in 
Section 3.1 (eq.{\ref{a1a1}). 

 How good is this approximation?
Keeping states for which $\kappa _{\vec{m}}\leq 0.5$ the 
correction terms ${\tilde{\Delta}_{i}}$ are $20\%$ of the 
field theory result for $T_{2}=100$
and this rises to $50\%$ for $T_{2}=10.$ The full threshold corrections
require the inclusion of the contribution above $\kappa _{\vec{m}}=0.5$ and
in the effective field theory approach these must be input as boundary
conditions for the RGE at the scale $\kappa _{\vec{m}}=0.5.$ Using the exact
string result we find that the combination of this contribution and the
correction terms ${\tilde{\Delta}_{i}}$ amount to roughly $50\%$ of the full
threshold correction for $T_{2}$ in the range $10$ to $100.$ The implication
of this is that while the effective field theory provides an excellent
description of the contribution of those states far below the string scale
it fails to give a very accurate determination of the full threshold
corrections because of the contribution of Kaluza Klein states close to the
string scale which have significant non-point-like coupling. We shall
discuss in Section \ref{hs} how to avoid this failing of the effective field
theory approach by improving the radiatively corrected calculation using the
full string threshold effects to determine the boundary conditions and to
take account of the non-pointlike coupling of those states close to the
string threshold. However, we note that at one loop order this uncertainty
only affects the determination of the string scale $M_{s}$. This follows
because, c.f. eq.(\ref{fullsum}), ${\tilde{\Delta}}_{i}$ is
proportional to 
${\overline{b}}_{i}$ and thus can be absorbed in a change of scale 
$M_{s}\rightarrow M_{s}^{\prime }$ in the first term. If we do this we can
extend $\kappa _{\vec{m}}$ to 1 and the ``field-theory'' term is exact at
one loop. Thus, using the effective field theory approach we lose the
accurate determination of the string scale, but otherwise the structure is
correct - it correctly reproduces the relative evolution of couplings and
the power-law behaviour at large radius (according to our discussion of
power-law running in Section \ref{eft}).

\subsection{Type I (I$^{\prime}$) string theory}

Threshold effects have also been calculated in Type I/I$^{\prime }$
string theories \cite{dudash,bachas,antoniadis_bachas,fabre}. These are of
much current interest because they allow (eq.(\ref{witten})) for a very low
string scale consistent with the 4D Planck mass. They may also accommodate
very large new dimensions in which the closed string states (gravitons,
etc.) propagate, giving rise to interesting new phenomena \cite{savas}.
Threshold effects in such models have quite a different character to those
of the weakly coupled heterotic string. The reasons are two-fold. Firstly
the states of the MSSM correspond to open string states with their quantum
numbers supplied by Chan-Paton charges. As a result the geometry associated
with one loop contribution involving the propagation of open string degrees
of freedom is now given by the annulus, ${\cal A}$, and the M\"{o}bius
strip, ${\cal M}$. This affects the way the ultraviolet cut-off appears, it
being determined by the lightest state in the crossed (closed-string) 
channel.
The second major difference is that in Type I$^\prime$
the open string has only winding modes (of mass $n M_s^2 R$, $n$
integer) with respect to ``off-the-brane'' dimensions 
(or Kaluza Klein modes  in Type I). As a result in the
case the compactification scale $R$ is larger than the string length scale,
these (winding) excitations are heavier than the string mass scale in 
Type I$^{\prime }$ theories. As we shall discuss, for the ${\cal N}=2$
contributions to threshold corrections, it is this scale and not the string
scale that acts as the ultraviolet cut-off. In this case one might worry
that an effective field theory (EFT) will not be able to describe such
radiative corrections coming from above the string scale. However, even in
this case, the EFT techniques work because the massive string states
fill ${\cal N}=4$ representations and do not contribute to the gauge coupling
evolution.

As we have already discussed, the corrections to gauge couplings come only
from the ${\cal N}=1$ (massless) sector and from the ${\cal N}=2$ massive
winding (or Kaluza Klein) sector. Let us consider the ${\cal N}=2$ sector
first. For Type I  strings at weak coupling, a result similar
to that of (\ref{gag}) has recently been computed in \cite{dudash}. The
massive ${\cal N}=2$ threshold 
corrections at $M_s$ have the form (for both ${\cal A}$
and ${\cal M}$ geometries) 
\begin{equation}  \label{type1boundary}
\Delta_a^{Type I}=\frac{1}{4\pi} \sum_{i}^{}b_{a i}^{{\cal N}=2}
\int_{0}^{\infty} \frac{dt}{t}\,e^{-\pi M_i^2 t}
\sum_{(m_{1_i},m_{2_i})} 
e^{-\frac{\pi t}{\sqrt{G_i} Im U_i}\vert m_{1_i}+U_i m_{2_i}\vert ^2}
\end{equation}
where $M_i$ is the mass of the lightest 
${\cal N}=2$ state in the KK tower 
$(M_i\ll 1/ \tilde R_1,  1/\tilde R_2)$. Evaluating this gives 
the following ${\cal N}=2$ correction at string scale\cite{dudash}
\begin{equation}  \label{typeI}
\Delta_a^{Type I}=-\frac{1}{4\pi} \sum_{i}b_{a i}^{{\cal N}=2} 
\ln\left[\sqrt{G_i} ImU_i \, M_s^2\, |\eta(U_i)|^4 \right]
\end{equation}
$G_i$ is the metric on the torus $T_i$.
For the simple case of a two-torus  the behaviour of the thresholds in
the limit  $Im U=\tilde R_1/\tilde R_2$  is of order one
($\sqrt G =\tilde R_1\tilde R_2$) is 
logarithmic, $\Delta_a\sim \ln(\tilde R_1 \tilde R_2)$, 
while if one radius is much
smaller than the other,  the thresholds are linearly divergent, 
$\Delta_a\sim \tilde R_2/\tilde R_1$ (``power-law''
running\footnote{$R$ and $\tilde R$ are related by a T duality 
transformation.}).

Once again, it is of interest to discuss the origin of these results in the
context of an effective field theory. In this case the integral over $t$
ranges from 0 to $\infty $ with a lower cut-off. The upper limit corresponds
to the IR region and is regulated by the mass of the lightest ${\cal N}=2$
state. The lower limit probes the UV in the open string channel. In order to
determine the ultraviolet cutoff imposed by the string it is instructive to
rewrite the threshold correction as an integral over $l$ in the crossed
(closed string) channel given by $l=1/t$ for ${\cal A}$ and $l=1/(4t)$
for 
${\cal M}$. As we discussed above, in the large radius limit it is necessary
to adopt the Type I$^{\prime }$ interpretation. Then 
\begin{eqnarray}
\Delta _{a}^{TypeI^{\prime }} &=&\frac{1}{4\pi }
\sum_{i}^{{}}b_{ai}^{{\cal N}=2}\int_{0}^{\infty }\frac{dt}{t}e^{-\pi
M_{i}^{2}t}\sum_{(m_{1_i},m_{2_i})} e^{-{\pi }t\left[
(m_{1_i} R_{1})^{2}+(m_{2_i} R_{2})^{2}\right] } \\
&=&\frac{1}{4\pi }\sum_{i}^{{}}\frac{b_{ai}^{{\cal N}=2}}{R_{1}R_{2}}
\int_{0}^{\infty }{dl}\,e^{-\pi M_{i}^{2}/l}
\sum_{(n_{1_i},n_{2_i})}
 e^{-\pi l \left[(n_{1_i}/R_{1})^{2}+(n_{2_i}/R_{2})^{2}\right] }  
\label{typeI_threshold}
\end{eqnarray}
where in deriving eq.(\ref{typeI_threshold}) we have performed a Poisson
resummation of the first term before changing variables. For inverse radii
roughly equal the integral over $l$ is cut-off for $l <1/R^{2}$ 
corresponding to $t\geq 1/\tilde R^{2}$ for ${\cal A}$ 
(or $t\geq 1/(4\tilde R^{2})$ for ${\cal M}$). 
Using this  eq.(\ref {typeI_threshold}) gives rise to 
the term $\propto \ln (R^{2})$ of eq.(\ref{typeI}). 
For the case the radii are hierarchically different, $R_{1}\gg R_{2}$,
states in  (\ref{typeI_threshold}) of mass $n/R_1$ 
contribute to the integral in (\ref{typeI_threshold}) which 
is now cut-off for $l<1/R_2$ (or for $t>1/\tilde R_2^2$).
The theory is effectively five dimensional and the couplings 
evolve as a linear power  $(R_{1}/R_{2})$ corresponding to the
large $Im U$ behaviour of eq.(\ref{typeI}).
In both cases only a finite number of states contribute allowing for an
effective field theory description of the region up to the cut-off scale.
Once again the EFT with a sharp cut-off does not accurately describe the
effect of the modes close to the cut-off scale, but as we saw in the
heterotic case, this just amounts to an uncertainty in the cut-off scale at
one loop order. The important feature that allows this interpretation, even
for the case the cut-off scale is much larger than the string scale, is the
decoupling of the massive string states.

We turn now to the contribution of the massless modes (before ``Higgsing'')
in the ${\cal N}=1$ sector. This has recently been calculated \cite{dudash}
in $Z_N$ orbifolds with $N$ a prime integer. These are the simplest 4D
models with ${\cal N}=1$ Supersymmetry having no 5-branes in the spectrum.
The important point to note is that the ${\cal N}=1$ sector does not include
any winding modes and is thus independent of the compactification radius. As
a result the regulated contribution is cut-off at the string scale $M_s$,
rather than the compactification scale.

The analysis of gauge coupling unification is somewhat complicated by the
presence of the twisted NS-NS moduli, $m_{k}$ the ``blowing-up'' modes of
the orbifold. These have non-universal couplings to the gauge fields and
this could affect the unification scale, \cite{louis1,louis2,louis3}. For
example, in the $Z_{3}$ model, a linear symmetric combination, $M$, of the
27 twisted moduli couples to the gauge fields. The gauge kinetic term has
the form 
\begin{equation}
f_{a}=S+S_{a}M
\end{equation}
where $S$ is the dilaton and $S_{a}$ are gauge group dependent factors.
These have been calculated \cite{dudash} and found to be proportional to the
one loop $\beta $ function coefficients, $b_{a}$. As a result if the vacuum
expectation value of $M$ is non-zero, it will lead to a shift in the
one-loop unification scale, $M_{s}\rightarrow M_{s}e^{<M>}$. The $Z_{3}$
model has a single anomalous $U(1)_{X}$ under which $M$ transforms in the
manner needed to cancel the anomalies via the Green-Schwarz mechanism.
Associated with this is the Fayet-Iliopoulos D-term which is sensitive
to 
$<M>$. Requiring that the non-Abelian gauge group remains unbroken, the
vanishing of the Fayet-Iliopoulos term forces $<M>$ to vanish \cite{dudash}.
A similar result is found for the other models studied. As 
a result\footnote{Provided higher order terms do not shift the minimum
of the potential for M.}
the unification scale of the massless sector remains the string scale,
$M_{s} $. In the case this scale is low, the only possible way to achieve
gauge unification is via the ${\cal N}=2$ correction of (\ref{typeI}). We
shall discuss this possibility in Section 6.

\section{Kaluza-Klein thresholds and RGE equations in 
field theory\label{eft}}

As we have just discussed, many features of the low energy structure of
string theories with large compactification radii can actually be described
in terms of a four dimensional effective field theory. In toroidal
compactification the existence of new compact spatial dimensions
requires 
$\Phi (x,y)=\Phi (x,y+2\pi R)$ where $\Phi (x,y)$ is a field, $x$ denotes the
four dimensional space-time coordinates and $y$ stands for the additional
(compactified) spatial dimensions assumed to be all of equal radius, $R$.
The coefficients (operators) $\Phi ^{(n)}(x)$ of the Fourier expanded
field 
$\Phi $ with respect to the new spatial coordinates $y$ represent the
Kaluza-Klein (KK) modes. As we increase the energy scale the new dimensions
will open up and this corresponds to the appearance in the spectrum of new
heavy states (excited modes). Their (bare) mass, $\mu _{\vec{m}}^{o},$ is
determined in terms of the inverse size of the extra (compact) spatial
dimensions , $\mu _{0}\equiv 1/R$ %\end{mathletters}
\begin{equation}
\mu _{\vec{m}}^{o2}=\mu _{0}^{2}(m_{1}^{2}+m_{2}^{2}+\cdots m_{\delta
}^{2})+m_{0}^{2}  \label{masskaluza}
\end{equation}
where $\delta $ is the number of additional dimensions, $\vec{m}\equiv
(m_{1},m_{2},\cdots ,m_{\delta })$, the integers $m_{i}$ are Kaluza-Klein
excitation numbers with integer values. The parameter $m_{0}$ denotes the
mass of ``zero-modes'' which we drop in this discussion, assuming it to be
much smaller than $\mu _{0}.$

Once the quantum numbers of the Kaluza Klein states are specified together
with their interactions with the light spectrum, one can apply the
renormalisation group evolution (RGE) to analyse the implications such a
spectrum has on the unification of the couplings. The use of the RGE
equations is justified in this context by the observation that an effective
field theory (below the string scale $M_s$) in $4+\delta $ dimensions can be
described by a renormalisable field theory with a {\it finite} number of
Kaluza Klein states below this scale\footnote{%
For a discussion on renormalisability, see Dienes et. al, 
\cite{dienes_v1}.}. 
This is just the structure we found in eq.(\ref{stringcutoff}) where a
finite number of states reproduced the usual field theory logarithmic
contributions. At and above the string scale it is necessary to include
string excitations and perform a full string calculation. To match the two
theories at the string scale we need to introduce string boundary conditions
for the RGE evolution below the string scale. We shall discuss the inclusion
of boundary conditions in Section \ref{hs}.

In this Section we determine the effects of the (finite number of)
Kaluza-Klein states on the running of the gauge couplings. Part of the
contribution to the running of the gauge couplings will be perturbatively
exact, while the contribution corresponding to the MSSM matter wave-function
renormalisation will be determined to two-loop order only. The RGE equation,
expressed in terms of the wave-function renormalisation coefficients, can be
derived from the ``NSVZ beta function'' \cite{shifman,NSVZ,murayama2}
(generalized in \cite{kaplunovsky} to local supersymmetric effective field
theories).

We shall calculate the evolution of the gauge couplings in a model with the
full MSSM spectrum together with the associated set of Kaluza-Klein
excitations. The latter may be associated with any combination of the gauge
sector, the Higgs fields and even with the fermionic sector, the choice
being determined by the particular model one chooses. In this section we
will allow for any of these possibilities, leaving the calculation in one
specific model to Section \ref{models}.

\subsection{Case 1: Kaluza Klein states with mass below string scale}

For generality, consider first the effect of a tower of Kaluza-Klein states
each state having $\Delta b_{i}=T(R^{i})$, $(i=1,2,3)$, as its contribution
to the one-loop $\beta$ function coefficients\footnote{
The Kaluza Klein states contributing to gauge coupling evolution fill
in ${\cal N}=2$ multiplets; here we use the ${\cal N}=1$ language.}. 
This tower
could be associated with any low-energy state, function of the value
of $T(R^{i})$. The number of the Kaluza-Klein states (including the 
zero-modes)
associated with any such low-energy state in representation $R^{i}$ is set
by the number of solutions (including the trivial one) \cite{dienes_v1} to
the equation which sets the upper limit to the mass of these excited modes,
corresponding to the cut-off in the effective field theory. This is given by 
\begin{equation}  \label{upperlimit}
0\leq (m_{1}^{2}+m_{2}^{2}+\cdots +m_{\delta }^{2})\leq \left[ \frac{\Lambda 
}{\mu _{0}}\right] ^{2}
\end{equation}
where $\Lambda $ is the high scale cut-off of the 4-dimensional effective
field theory and $\mu _{0}$ is the compactification scale, equal to the
inverse of the radius of the extra spatial dimensions (we assume that all
additional spatial dimensions have equal radius, thus the mass splitting of
KK states is proportional to $\mu_0$). The number of states is approximated
by the volume of a $\delta $-dimensional sphere \cite{dienes_v1} of
radius 
$\Lambda /\mu _{0}$, given by 
\begin{equation}
N(\Lambda ,\mu _{0};\delta )\cong \frac{\pi ^{\delta /2}}{\Gamma (1+\delta
/2)}\left[ \frac{\Lambda }{\mu _{0}}\right] ^{\delta }  \label{number}
\end{equation}
The ``radial'' degeneracy $\sigma $ of a level of fixed energy (i.e. a fixed
value for $\mu _{\vec{m}}^{o2}=\mu _{0}^{2}(m_{1}^{2}+m_{2}^{2}+\cdots
+m_{\delta }^{2})$) is therefore given by 
\begin{equation}
\sigma (\mu _{\vec{m}}^{o},\mu _{0};\delta )\cong \frac{dN}{d\frac{\Lambda 
} {\mu _{0}}}\bigg\vert_{\Lambda =\mu _{\vec{m}}^{o}}=\frac{\delta \pi
^{\delta /2}}{\Gamma (1+\delta /2)}
\left[ \frac{\mu _{\vec{m}}^{o}}{\mu _{0}}
\right] ^{\delta -1}  \label{density}
\end{equation}
which increases with the scale. This means that the Kaluza-Klein states will
be ``turned on'' in an increasing number at a higher scale and with a strong
effect on the RGE equations within a small range of energy. 
Eqs (\ref{number}), 
(\ref{density}) will prove useful in evaluating the contribution of the
tower states to the RGE equations.

The RGE equations for the gauge couplings \cite{shifman,NSVZ,murayama2}
include contributions from the gauge wave-function renormalisation and
matter wave-function renormalisation, which can be easily 
obtained\footnote{See also \cite{zurab3} for a discussion of 
radiative corrections with Kaluza
Klein towers.} by formally integrating the {\it exact} NSVZ beta 
function\footnote{Beyond two-loop order, 
this is regularisation scheme dependent \cite{jones2}.} 
\cite{shifman,NSVZ,murayama2} 
\begin{equation}
\beta (\alpha_i )^{NSVZ}\equiv \frac{d\alpha_i }{d(\ln Q )}=-\frac{
\alpha_i^{2}} {2\pi }\left[ 3T_i(G)-\sum_{\psi }^{{}}T(R^i_{\psi })(1-
2\gamma _{\psi })\right] \left( {1-T_i(G)\frac{\alpha_i}{2\pi }}\right) ^{-1}
\label{shifmanbetaalpha}
\end{equation}
with the definition ($Q$ is the scale) 
\[
\gamma _{\psi }=-\frac{1}{4\pi} \frac{d\ln Z_{\psi }}{dt},\;\;\;\;\;\; t= 
\frac{1}{2 \pi}\ln Q 
\]
and where $T_i(G)$ and $T(R^i_{\psi })$ represent the Dynkin index for the
adjoint representation and for $R^i_{\psi }$ representation respectively
(not necessarily the fundamental one) which contribute to the running of $%
\alpha_i$, $i=\{1,2,3\}$. The above sum runs over {\it all} matter fields $%
\psi $ in the representation $R^i_\psi$ and this includes the high energy
(Kaluza Klein) and the low energy (MSSM) spectrum. Eq.(\ref{shifmanbetaalpha}%
) can be re-written as follows 
\begin{equation}  \label{shf}
-\frac{d \alpha_i^{-1}}{d \ln Q} =\frac{1}{2\pi}\left[ T_i(G) \frac{d
\ln\alpha_i}{d \ln Q} -3 T_i(G)+ \sum_{\psi}T(R^i_{\psi})(1-2\gamma_{\psi}) 
\right]
\end{equation}
One can integrate this equation\footnote{In fact 
it is the NSVZ function which is derived from an equation \cite
{shifman,murayama2} recovered here by integrating back (\ref
{shifmanbetaalpha}).} \cite{shifman} to give the exact contributions to all
orders in perturbation theory to the running of the gauge couplings for the
particular spectrum considered. The results depend however on the
wave-function renormalisation coefficients which are not known to all orders
and this makes the results less useful. Still, for the effects of the heavy
states to $\alpha_i^{-1}$ one finds exact results \cite{shifman,murayama2}
which depend on the {\it bare} mass of the states and not on the factors $Z$
and this is useful for our phenomenological predictions.

More explicitly, consider the Kaluza-Klein tower of states of bare
mass  $\mu_{\vec{m}}^{o}$ (eq.(\ref{masskaluza})) associated with a {\it particular}
low-energy state in representation $R^{i}$, $i=1,2,3$, with Dynkin index $%
T(R^{i})\equiv \Delta b_{i}$. In the RGE they decouple at a scale equal to
their {\it physical} mass, $\mu _{\vec{m}}$ \cite{murayama2}. After
performing an integration of eq.(\ref{shf}), we obtain the generic
contribution of this tower of states to the RGE equations for $\alpha
_{i}^{-1}(\mu _{0})$ 
\begin{eqnarray}
{\cal A}_{i} &=&\frac{\Delta b_{i}}{2\pi }\sum_{\psi }^{{}}\int_{\mu
_{0}}^{\Lambda }\;(1-2\,\gamma _{\psi })d\ln \mu   \nonumber \\
&=&\frac{\Delta b_{i}}{2\pi }\sum_{\vec{m}}^{{|\vec{m}|\leq \Lambda /\mu _{0}%
}}\sigma _{{\vec{m}}}\ln \frac{\Lambda Z(\Lambda )}{\mu _{{\vec{m}}}Z_{\vec{m%
}}(\mu _{{\vec{m}}})}=\frac{\Delta b_{i}}{2\pi }\sum_{\vec{m}}^{{|\vec{m}%
|\leq \Lambda /\mu _{0}}}\sigma _{{\vec{m}}}\ln \frac{\Lambda }{\mu _{\vec{m}%
}^{o}}  \label{a1a1}
\end{eqnarray}
In (\ref{a1a1}) $\sigma _{{\vec{m}}}$ is the ``radial'' degeneracy which can
be approximated by eq.(\ref{density}). The degeneracy accounts for   all
possible configurations ${\vec{m}}\equiv (m_{1},m_{2},\cdots ,m_{\delta })$
with {\it fixed} value for $\mu _{{\vec{m}}}^{o2}\equiv \mu
_{0}^{2}(m_{1}^{2}+m_{2}^{2}+\cdots +m_{\delta }^{2})$. In eq.(\ref{a1a1})
we have also used the mass renormalisation equation 
\begin{equation}
\mu _{\vec{m}}Z_{\vec{m}}(\mu _{\vec{m}})=\mu _{\vec{m}}^{o}Z(\Lambda )
\end{equation}
The mass renormalisation equation is exact to all orders in perturbation
theory \cite{shifman,NSVZ} and gives a result (eq.(\ref{a1a1})) depending on
the {\it bare} mass of these states only. This mechanism is general, as long
as Supersymmetry is present \cite{shifman,NSVZ} and shows how to sum up the
heavy modes' contribution to the RGE equations. Strictly speaking the
mechanism applies only to ${\cal N}=1$ Kaluza Klein states (i.e. zero modes)
while for ${\cal N}=2$ states we have\footnote{%
One could simply use \cite{murayama2} for the RGE evolution, which is just
the integral of NSVZ function. The ${\cal N}=2$ Kaluza Klein states (in
addition to the MSSM) which in ${\cal N}=2$ language do not have
wavefunction renormalisation, give only an overall contribution to $1/\alpha
_{i}$ equal to ${\overline{b}}_{i}/(2\pi )\sum \ln \Lambda /\mu _{\vec{m}%
}^{o}$ with ${\overline{b}}_{i}$ accounting for all ${\cal N}=2$ Kaluza
Klein (gauge bosons and/or matter fields) contributions. For a
self-contained approach and explicit presentation of the mass
renormalisation of the ${\cal N}=1$ components we used here the
``NSVZ'' $\beta $ function and ${\cal N}=1$ language, with the same result for
the RGE.}  $Z_{\vec{m}}=1$.

Note that the result of eq.(\ref{a1a1}) is exactly the string theory term of
eq.(\ref{stringcutoff}) (with $\delta =2$ since the string calculation above
is valid for a two torus) provided that the Kaluza Klein spectrum is the
same in both cases. The interpretation of the scale $\Lambda $ is obvious
from eq.(\ref{stringcutoff}). It shows that the string and field theory
results for the Kaluza Klein thresholds are equal for the sum over those
states whose mass satisfies 
\begin{equation}
\mu _{\vec{m}}^{o}\equiv \mu _{0}|{\vec{m}}|\ll M_{s}
\end{equation}
In terms of the RGE evaluation of these terms what is important is the
logarithmic $\mu _{\vec{m}}^{o}$ dependence of the result and one may
readily check that this is given by eq(\ref{stringcutoff}) to an accuracy of 
$10\%$ for $\mu _{\vec{m}}^{o}\leq M_{s}/10.$ This is equivalent to using
the RGE up to the   scale $\Lambda =M_{s}/10.$ Above this scale the field
theory estimate of the Kaluza Klein contribution deviates from the string
result because of the pointlike coupling assumed in the effective field
theory result. The effect of the states in this region must be included as
boundary conditions for the RGE evolution using the full string theory
calculation. This is the approach we take in Section \ref{hs} to show how to
determine the radiative corrections using the string theory threshold
corrections directly.

The result of eq.(\ref{a1a1}) can also be written as 
\begin{equation}  \label{ttt}
{\cal A}_{i}=\frac{\Delta b_{i}}{2\pi }\left[ N(\Lambda ,\mu _{0};\delta )-1 %
\right] \;\ln \frac{\Lambda }{\mu _{0}}-\frac{\Delta b_{i}}{2\pi }\sum_{{\ 
\vec{m}}}^{{|\vec{m}|\leq \Lambda/\mu_0 }} \sigma _{{\vec{m}}}\ln \frac{\mu
_{{\vec{m}}}^{o}}{\mu _{0}}
\end{equation}
The sum $\sum_{\vec{m}}^{{}}\sigma _{\vec{m}}$ was replaced by $N(\Lambda
,\mu _{0},\delta )-1$ which accounts for the number of {\it excited} modes
only, and this excludes the configuration of ``zero modes'' ${\vec{m}}%
=(0,0,\cdots,0)$. To evaluate the last sum of eq.(\ref{ttt}) we can {\it %
approximate} it by an integral with upper mass limit set by eq.(\ref
{upperlimit}) while the lower limit is the mass of the lowest excited mode, $%
|{\vec m}|=1$. The result obtained for the sum of eq.(\ref{ttt}) is then
given by 
\begin{eqnarray}  \label{tttt}
\sum_{{\vec m}}^{|\vec{m}|\leq \Lambda/\mu_0} \sigma_{{\vec m}} \ln \frac{%
\mu_{{\vec m}}^o}{\mu_0} &\cong& \frac{\delta\; \pi^{\delta/2}}{%
\Gamma(1+\delta/2)} \int_{1}^{\Lambda/\mu_0} z^{\delta-1} \ln z\; dz 
\nonumber \\
&=& N(\Lambda,\mu_0; \delta) \ln\frac{\Lambda}{\mu_0} -\frac{\pi^{\delta/2}}{
\delta\;\Gamma(1+\delta/2)} \left[\left(\frac{\Lambda}{\mu_0}
\right)^{\delta}-1\right]
\end{eqnarray}
Using this, one finds 
\begin{equation}
{\cal A}_i\cong\frac{\Delta b_i}{2\pi} \left\{ \frac{\pi^{\delta/2}}{
\delta\;\Gamma(1+\delta/2)} \left[\left(\frac{\Lambda}{\mu_0}
\right)^{\delta}-1\right]- \ln\frac{\Lambda}{\mu_0}\right\}  \label{a1}
\end{equation}
where ${\cal A}_i$ represents the {\it exact} contribution, to all orders in
perturbation theory of a tower of Kaluza-Klein states to the running of the
gauge couplings. This is the result for a tower of states associated with a
low-energy state in representation $R^i$, with the one-loop beta function
contribution $\Delta b_i=T(R^i)$. Equation (\ref{a1}) also takes account of
the heavy threshold effects. The power-law behaviour of eq.(\ref{a1}) is a
just a consequence of the (large) number of states we are summing over, each
of them giving an individual contribution to the RGE equations of {\it %
logarithmic} type. The result is similar to the ``one-loop'' result of \cite
{dienes_v1} obtained in the ``standard'' way. Note that the computation of
the last sum in equation (\ref{ttt}) is an approximation which works well
for large number of states and also uses the approximation (\ref{number})
which for $\delta=1,2,3$ is indeed reliable\footnote{%
For this compare the number of Kaluza Klein states of Table 3 of \cite
{bogdan} with that given by eq.(\ref{number}) for the ratio $M_s/\mu_0$
given in Table 3 of \cite{bogdan}.}. One should however ensure that this is
true for all phenomenological cases, particularly when the number of
Kaluza-Klein states is relatively low when eq.(\ref{number}) may not be
reliable. This approximation will be eliminated in  Section 4 (for the
case of weakly coupled heterotic string) were we match the field theory 
results with those from  string theory. This concludes the
calculation of the generic contribution of a tower of Kaluza-Klein states
associated with a low-energy state of Dynkin index $T(R^i)$.

For the general case the coefficient $\Delta b_i=T(R^i)$ in eq.(\ref{a1})
should be replaced by the total contribution to the one loop beta function
of the massive KK states considered, which we call\footnote{%
Kaluza Klein states are ${\cal N}=2$ multiplets hence ${\overline b}_i$ is
the ${\cal N}=2$ $\beta$ function, ${\overline b}_i=2 \sum_{\psi}^{}
T_i(\psi)-2 T_i(G)$.} ${\overline b_i}$. Then the gauge couplings at the
compactification scale have the generic form 
\begin{eqnarray}  \label{AA}
\alpha_i^{-1}(\mu_0)&=& \alpha_i^{-1}(\Lambda)+ \frac{\overline b_i}{2\pi}
\sum_{\vec m}^{|\vec{m}|\leq \Lambda/\mu_0} \sigma_{{\vec m}} \ln \frac{%
\Lambda}{\mu^o_{\vec m}}+l.s.~~contribution \\
&\cong &\alpha_i^{-1}(\Lambda)+\frac{{\overline b_i}}{2\pi}\left\{ \frac{
\pi^{\delta/2}}{\delta\;\Gamma(1+\delta/2)} \left[\left(\frac{\Lambda}{\mu_0}
\right)^{\delta}-1\right]- \ln\frac{\Lambda}{\mu_0}\right
\}+l.s.~~contribution  \label{BB}
\end{eqnarray}
where the power-law term is perturbatively exact. On the other hand, the
last term due to the {\bf l}ight {\bf s}tates (l.s.) (such as those of the
MSSM) can only be computed perturbatively up to some given order. This
concludes the calculation of the {\it explicit} contribution of the heavy
Kaluza-Klein states to the RGE equations.

We now evaluate the contribution of the matter fields of the MSSM to the
running of the gauge couplings, in two-loop order. For this we need to
evaluate the anomalous dimensions of the MSSM matter fields (which we call $%
\gamma_\phi$) in one-loop order only \cite{shifman,murayama2}. These
quantities evaluated above the decompactification scale $\mu_0$ are changed
from their usual MSSM values due to the presence of the tower of
Kaluza-Klein states. In this way Kaluza Klein states give a second, {\it %
indirect} contribution, of two-loop (and higher) order to the RGE equations.
To obtain a definite expression for the value of the anomalous dimensions of
the MSSM matter fields one needs to determine further model-dependent
properties i.e. the interaction Lagrangian between the low-energy fields and
the excited Kaluza-Klein states which can contribute to the wavefunction
renormalisation of the MSSM states. To keep a model independent approach, we
separate the contributions to $\gamma_\phi$ in two terms, one due to the
MSSM states alone (denoted by $\gamma_{\phi}^o$) and one including Kaluza
Klein states, which we call $\Delta\gamma_\phi$. Therefore 
\begin{equation}  \label{sumg}
\gamma_\phi=\gamma^o_\phi+\Delta\gamma_\phi
\end{equation}
If we ignore any Yukawa contribution\footnote{%
Yukawa effects on the light spectrum wavefunction renormalisation will be
neglected throughout this paper; they can be included in (\ref{sumg}) as
well, as in \cite{dgl}.} the value of $\gamma^o_\phi$ has the following
expression, valid between the scale $M_z$ and $\Lambda$ 
\begin{equation}  \label{gamma0}
4 \pi \gamma^o_{\phi}(Q) = - \sum_{k=1}^{3} 2 \alpha_k C_k(\phi)+{\cal O}%
(\alpha^2) \;\;\;\;\;\;\;\;\; M_z < Q < \Lambda
\end{equation}
where $\phi$ can stand for the chiral superfields of the MSSM and $C_k(\phi)$
represents the quadratic Casimir operator. We therefore find the following
contribution to $\alpha_i^{-1}(\mu_0)$ from the MSSM fields $\phi_j$ (three
families, j=1,2,3) and the two Higgs doublets 
\begin{eqnarray}
{\cal B}_{i}&\equiv&\sum_{j=1}^{3}\sum_{\phi_j}^{} \int_{\mu_0}^{\Lambda} 
\frac{d \ln \mu}{2 \pi}\; T(R^i_{\phi_j})\; (1-2 \gamma_{\phi_j})
+\sum_{a=u,d}^{3} \int_{\mu_0}^{\Lambda}\frac{d \ln \mu}{2 \pi}\;
T(R^i_{H_a})\; (1-2 \gamma_{H_a})  \nonumber \\
&=&\frac{b_i+3 T_i(G)}{2\pi}\ln\frac{\Lambda}{\mu_0} +\frac{1}{4\pi}
\left(b_{ik}-2 \; b_k T_k(G) \delta_{ik} \right)\int_{\mu_0}^{\Lambda}\frac{
d \ln \mu}{2 \pi}\; \alpha_k(\mu)  \nonumber \\
&-& \sum_{j=1}^{3}\sum_{\phi_j}^{} T(R^i_{\phi_j}) \int_{\mu_0}^{\Lambda} 
\frac{d \ln \mu}{2 \pi}\; 2\; \Delta \gamma_{\phi_j}-\sum_{a=u,d}^{}
T(R^i_{H_a}) \int_{\mu_0}^{\Lambda}\frac{d \ln \mu}{2 \pi}\; 2\; \Delta
\gamma_{H_a}  \label{a2}
\end{eqnarray}
where $(b_i+3 T_i(G))$ is the MSSM {\it matter} contribution, also $%
b_i=(33/5,1,-3)$, $b_{ik}$ is the two-loop MSSM beta function, and $%
T_k(G)=\{0,2,3\}_k$ is the Dynkin index for the adjoint representation of
the gauge group G. Since in eq.(\ref{a2}) we used one-loop order values for
the anomalous dimensions, this expression will be valid in {\it two-loop}
order only.

We now add together all the contributions to the RGE equations for the gauge
couplings which we evaluate just below the scale $\mu_0$, in two loop order.
This includes the contributions of eq.(\ref{a1}) modified by the factor ${\
\overline b}_i$ of (\ref{BB}) and the contributions of eq.(\ref{a2}) as well
as those due to the wave-function renormalisation of the MSSM gauge bosons.
The latter is actually exact to all orders in perturbation theory \cite
{shifman} and is equal to ${T_i(G)}/({2\pi}) \ln({\alpha_{\Lambda}}/{%
\alpha_i(\mu_0)})$. We thus obtain the following result (using (\ref{shf})) 
\begin{eqnarray}
\alpha_i^{-1}(\mu_0) & \cong &\alpha^{-1}_i(\Lambda) +\frac{\overline{b}_i}{%
2 \pi} \left\{\frac{\pi^{\delta/2}}{\delta\;\Gamma(1+\delta/2)} \left[\left(%
\frac{ \Lambda}{\mu_0} \right)^{\delta}-1\right]-\ln\frac{\Lambda}{\mu_0}%
\right\} + \frac{b_i}{2\pi}\ln\frac{\Lambda}{\mu_0} +\frac{T_i(G)}{2\pi} \ln%
\frac{\alpha_i(\Lambda)}{\alpha_i(\mu_0)}  \nonumber \\
&+& \frac{1}{4\pi}\sum_{k=1}^{3}\left(b_{ik}-2\; b_k\;
T_k(G)\delta_{ik}\right) \int_{\mu_0}^{\Lambda} \frac{d \ln\mu}{2\pi}
\;\alpha_k(\mu)  \nonumber \\
&-& \sum_{j=1}^{3}\sum_{\phi_j}^{} T(R^i_{\phi_j}) \int_{\mu_0}^{\Lambda} 
\frac{d \ln \mu}{2 \pi}\; 2\; \Delta \gamma_{\phi_j} -\sum_{a=u,d}^{}
T(R^i_{H_a}) \int_{\mu_0}^{\Lambda} \frac{d \ln \mu}{2 \pi}\;2 \; \Delta
\gamma_{H_a}  \label{RGEdecomp}
\end{eqnarray}
where $\overline b_i$ is the one loop contribution to beta function due to
those (light) states having Kaluza-Klein towers. Beyond the scale $\Lambda$
a more fundamental theory (of strings) is supposed to be valid. From the
scale $\mu_0$ down to $M_z$ scale, only the usual MSSM running is supposed
to take place and it affects the running of the gauge couplings in a manner
presented in \cite{extradim}.

Equation (\ref{RGEdecomp}) provides the general result for the running of
the gauge couplings in the presence of Kaluza-Klein states, derived on
purely field theoretical grounds. The symbol ``$\cong $'' accounts for the
approximation (\ref{tttt}) of the discrete sum over the KK thresholds by an
integral which gives the power-law term in (\ref{RGEdecomp}). Although this
is a good approximation, it may be avoided by doing the sum explicitly
as we show it for the case of weakly coupled heterotic string in 
Section \ref{hs}.  In the limit $\delta \rightarrow 0$ one recovers the
standard RGE of the MSSM.

To go further we need to know something about $\Delta\gamma$'s. Let us keep
only gauge contributions to $\Delta\gamma$ and work with the
 weakly coupled heterotic string. 
For the orbifold compactification procedure we adopt, we
assume that the Higgs and MSSM fields are situated at the fixed points of
the orbifold and thus do not have Kaluza Klein excitations. In orbifold
compactification the gauge coupling between two massless twisted modes (i.e.
a low-energy state without KK modes) situated at the {\it same} fixed point
of the orbifold and the (untwisted) gauge excitations of level $\vec{m}$ is
modified \cite{vafa,antoniadis} from the ordinary gauge coupling $\alpha
_{j} $, giving 
\begin{equation}
4\pi \Delta \gamma _{\phi }(Q)=-f_{\phi }(Q,\mu _{0};\delta
)\sum_{k=1}^{3}2\alpha _{k}(Q)C_{k}(\phi )\;\;\;\;\;\;\;\;\mu _{0}\leq Q\leq
\Lambda  \label{deltagamma}
\end{equation}
with 
\begin{equation}  \label{factors}
f(Q,\mu _{0},\delta )=\sum_{\vec m}^{|{\vec m}| \le Q/\mu_0} \sigma (
\mu_{\vec m}, \mu_0,\delta) \,\rho ^{-\frac{|{\vec{m}}|^{2}\mu _{0}^{2}}{%
M_{s}^{2}}}  \label{formfactor}
\end{equation}
where $\rho $ is a number which depends on the orbifold twist \cite{vafa},
of order 10 or larger. Here $\sigma (\mu_{\vec m},\mu_0,\delta)$ is the
degeneracy of Kaluza-Klein states coupling to the two massless twisted modes
and accounts for the sum of all effects due to excited KK modes of fixed $|%
\vec{m}|$. We see that the field theory result is again obtained in the
limit $\mu _{\vec{m}}^{o 2}\equiv |{\vec{m}}|^{2}\mu_{0}^{2}\ll M_{s}^{2}$
when the orbifold information ($\rho $ dependence in (\ref{factors})) is
lost and $f$ becomes equal to a {\it finite} number of excited Kaluza Klein
gauge bosons, equal to $N(Q,\mu_0,\delta)-1$ with $Q$ restricted to values
below the string scale.

Since we are keeping only gauge interactions, the three generations of the
MSSM have all the same function $f_{\phi }(Q,\mu _{0};\delta )$ and with the
corresponding function for the Higgs fields which we denote by $g(\mu ,\mu
_{0};\delta )$, we obtain the final form for the running of the gauge
couplings (with $\alpha _{i}(\Lambda )\rightarrow \alpha _{\Lambda }$) 
\begin{eqnarray}
\alpha _{i}^{-1}(\mu _{0}) &\cong &\alpha _{\Lambda }^{-1}+\frac{\overline{b}%
_{i}}{2\pi }\left\{ \frac{\pi ^{\delta /2}}{\delta \;\Gamma (1+\delta /2)}%
\left[ \left( \frac{\Lambda }{\mu _{0}}\right) ^{\delta }-1\right] -\ln 
\frac{\Lambda }{\mu _{0}}\right\} +\frac{b_{i}}{2\pi }\ln \frac{\Lambda }{%
\mu _{0}}+\frac{T_{i}(G)}{2\pi }\ln \frac{\alpha _{\Lambda }}{\alpha
_{i}(\mu _{0})}  \nonumber \\
&+&\frac{1}{4\pi }\sum_{k=1}^{3}\left( b_{ik}-2\;b_{k}\;T_{k}(G)\delta
_{ik}\right) \int_{\mu _{0}}^{\Lambda }\frac{d\ln \mu }{2\pi }\;\alpha
_{k}(\mu )  \nonumber \\
&+&\frac{1}{4\pi }\sum_{k=1}^{3}\left( b_{ik}-2\;b_{k}\;T_{k}(G)\delta
_{ik}-b_{ik}^{H}\right) \int_{\mu _{0}}^{\Lambda }\frac{d\ln \mu }{2\pi }%
\;\alpha _{k}(\mu )\;f(\mu ,\mu _{0};\delta )  \nonumber \\
&+&\frac{1}{4\pi }\sum_{k=1}^{3}b_{ik}^{H}\int_{\mu _{0}}^{\Lambda }\frac{%
d\ln \mu }{2\pi }\;\alpha _{k}(\mu )\;g(\mu ,\mu _{0};\delta )
\label{RGEfinal}
\end{eqnarray}
For the present case when there are only Kaluza Klein modes for the gauge
bosons, $f$ and $g$ are equal (and given by the number of Kaluza Klein
excitations of the gauge bosons) in (\ref{RGEfinal}). For the general case
when the Higgs fields and (or) MSSM fermions have KK excitations as well,
then $g$ and (or) $f$ are equal to zero as the states in the loop which
generates $\Delta \gamma _{\phi ,H}$ belong to complete ${\cal N}=2$
multiplets. 

The term with the coefficient $b_{ik}^{H}$ accounts for two loop 
effects to 
$\alpha _{i}^{-1}$ due to the MSSM Higgs fields through their {\it one-loop}
wavefunction renormalisation (due to {\it excited} KK modes in one line of
the loop\footnote{%
The corresponding Feynman diagram of this contribution is similar to that
leading to eq.(\ref{gamma0}).}) while the term $(b_{ik}-2b_{k}\;T_{k}(G)%
\delta _{ik}-b_{ik}^{H})$ is the equivalent contribution of the one-loop
wavefunction renormalisation of the three MSSM families (due to {\it excited}
KK modes). The term with the coefficient $(b_{ik}-2b_{k}T_{k}\delta _{ik})$
accounts for the ``standard'' two-loop MSSM effects to $\alpha _{i}^{-1}$
due to the wavefunction renormalisation of the three MSSM generations and
Higgs fields. The one-loop gauge wavefunction renormalisation effects are
accounted for by the term ${T^{i}(G)}/({2\pi })\ln ({\alpha _{\Lambda }}/{%
\alpha _{i}(\mu _{0})})$. Since the Kaluza-Klein states belong to ${\cal N}=2
$ supermultiplets this term is perturbatively exact. Note that the departure
from point-like coupling observed at one-loop level occurs at two loops as
well due to the presence of the functions $f$,$g$. For $\Lambda $ well below
the string scale two loop terms may be reliably calculated using point-like
couplings. The modification due to the non-pointlike coupling of states
close to the string scale is a relatively small effect. Finally, due to the
power-law behaviour of the couplings, the familiar two-loop running $%
\thicksim \log (\alpha (\Lambda )/\alpha (Q))$ of the MSSM is replaced by a
leading dependence of the type $\log (\Lambda /m)$ characteristic of the
one-loop terms in the MSSM. This can be seen by inserting the one-loop
power-like running for the gauge coupling in the integrals of eq.(\ref
{RGEfinal})

\subsubsection{Convergence of the perturbative expansion\label{pe}}

We are now in a position to determine whether the perturbative expansion
converges in the presence of the massive tower of Kaluza Klein states. As we
have noted the contribution involving only Kaluza Klein modes is
perturbatively exact. However the term corresponding to the wave function
renormalisation of the light states has contributions to all orders in
perturbation theory. From eq(\ref{RGEfinal}) we see that the two loop term
is of $O(\alpha N/4\pi )$ where $N$ counts the number of Kaluza Klein
states. This is to be compared with the one loop contribution which occurs
at $O(N)$ so the two loop term is suppressed at $O(\alpha )$ with respect to
the one loop term, just as in the case without Kaluza Klein towers. However
at higher order we expect terms of $O((\alpha N/4\pi )^{m})$ due to the
contribution of Kaluza Klein modes to the wavefunction renormalisation of
the MSSM spectrum, see eq.(\ref{deltagamma}) where higher order
contributions have been neglected. This means that these terms are
suppressed with respect to the {\it two loop} terms only if $N\alpha \ll 
{\cal O}(4\pi )$. As we will see in  a specific example (Section 
\ref{models}), this condition may be
satisfied; if not corrections to $\alpha $ at $O(\alpha ^{2})$ will have an
undetermined 
coefficient\footnote{If we reach the limit $N\alpha 
\approx {\cal O}(4\pi )$, a resummed ${\cal O}%
(1/N)$ perturbation series for anomalous dimensions of the light 
fields, eq.(\ref{sumg}) should be used \cite{jones} although 
its small radius of
convergence ($N\alpha <{\cal O}(6\pi )$) brings little improvement.
Beyond this limit we reach the non-perturbative regime.}.

\subsection{Case 2: Type I$^{\prime}$ strings with $M_s < M_{Winding}$}

For the case of Type I$^{\prime}$ strings with the winding mode scale {\it %
above} the string scale the situation is simpler. Below the string scale we
have standard MSSM running. Above the string scale only ${\cal N}=2$ massive
multiplets contribute at one loop order and the contribution is cut-off near
the first winding mode excitation \cite{bachas}. 
As discussed above, this contribution is perturbatively exact.
Its implications are discussed in Section \ref{type1}.

\section{RGE equations with full (heterotic) string thresholds\label{hs}}

As we discussed in Section \ref{string} the sum of Kaluza Klein modes with
mass less than or equal to the string scale accurately gives the string
threshold corrections even for moderate compactification radii, as small as
only a factor of 2 larger than the string scale. However the field theory
approximation fails to generate the correct contribution of the Kaluza Klein
modes with mass close to the string scale because it assumes a pointlike
coupling for all states. It was for this reason we argued in Section \ref
{eft} that the cutoff for the effective field theory had to be set somewhat
lower than the string scale but in this case significant corrections arise
from the need to impose boundary conditions at this scale. These boundary
conditions are determined by the effect of the states lying between the
cutoff scale and the string scale. It is possible to improve on this
situation in a given string theory by including the full string theory
threshold effects. We shall do this here for the case of the orbifold
compactification of the weakly coupled heterotic string but the method
immediately generalizes to the case of Type I strings or any other case in
which one can calculate the one-loop threshold effects. The full one-loop
string threshold effects are simply included in the general form for the RGE
equations by substituting the full string term (\ref{di}) in eq.(\ref
{RGEfinal}) rather than the field theory result eq.(\ref{a1a1}) or its
approximation (\ref{a1}). Doing this equation (\ref{RGEfinal}) becomes (with 
$\Lambda \rightarrow M_{s}$, $\alpha _{\Lambda }\rightarrow \alpha
_{s}\equiv \alpha (M_{s})$) 
\begin{eqnarray}
\alpha _{i}^{-1}(\mu _{0}) &=&\alpha _{s}^{-1}-\frac{\overline{b}_{i}}{4\pi }%
\ln \left\{ 4\pi ^{2}\,{|}\eta (i){|}^{4}\,\left( \frac{M_{s}}{\mu _{0}}%
\right) ^{2}\,{\bigg\vert}\eta \left[ \frac{i\,3\sqrt{3}\pi }{2e^{1-\gamma }}%
\,\left( \frac{M_{s}}{\mu _{0}}\right) ^{2}\right] {\bigg\vert}^{4}\right\} 
\nonumber \\
&+&\frac{b_{i}}{2\pi }\ln \frac{M_{s}}{\mu _{0}}+\frac{T_{i}(G)}{2\pi }\ln 
\frac{\alpha _{s}}{\alpha _{i}(\mu _{0})}  \nonumber \\
&+&\frac{1}{4\pi }\sum_{k=1}^{3}\left( b_{ik}-2\;b_{k}\;T_{k}(G)\delta
_{ik}\right) \int_{\mu _{0}}^{M_{s}}\frac{d\ln \mu }{2\pi }\;\alpha _{k}(\mu
)  \nonumber \\
&+&\frac{1}{4\pi }\sum_{k=1}^{3}\left( b_{ik}-2\;b_{k}\;T_{k}(G)\delta
_{ik}-b_{ik}^{H}\right) \int_{\mu _{0}}^{M_{s}}\frac{d\ln \mu }{2\pi }%
\;\alpha _{k}(\mu )\;f(\mu ,\mu _{0};\delta )  \nonumber \\
&+&\frac{1}{4\pi }\sum_{k=1}^{3}b_{ik}^{H}\int_{\mu _{0}}^{M_{s}}\frac{d\ln
\mu }{2\pi }\;\alpha _{k}(\mu )\;g(\mu ,\mu _{0};\delta )  \label{RGEstring}
\end{eqnarray}
This is an integral equation for $\alpha _{i}(\mu _{0})$ and can be solved
numerically with $\mu_0$ kept relatively close\footnote{See next
section for an example.} to $M_s$ 
to preserve the weakly coupled regime of the theory.
 Below the compactification scale $\mu _{0}$
only the MSSM spectrum and RGE ``running'' applies (see for example \cite
{extradim} for its explicit form). The appropriate matching field
theory-string theory results should therefore be done at the
compactification scale, $\mu _{0}$.

The use of the string boundary conditions eliminates the discrepancy 
\footnote{%
of up to $50\%$, see discussion at the end of Section 2.1} between the one
loop terms evaluated on pure field theory grounds and the exact string
result, which originated in the point-like nature of the couplings in field
theory. Regarding the two loop terms which we computed using a mixed field
theory-string theory formalism, their contribution relative to one loop
terms is in general small since they are suppressed by an ${\cal O}(\alpha )$
(see next section for an example). This contribution is about $4\%$ for the
prediction for the strong coupling at $M_{z}$ and thus the residual effect
of the discrepancy is suppressed at two loop level below $4\%\times 50\%=2\%$
(for a conservative estimate of the discrepancy between the field theory and
string result of $50\%$). This ``hybrid'' construction for two loop terms
and their small effect enable us to argue that eq.(\ref{RGEstring}) is close
to a full two-loop string result for the effects of string (Kaluza Klein and
winding) thresholds on the gauge couplings.

\section{A numerical example\label{models}}

We present for illustration of the methods developed here a simple model and
investigate its predictions following from the RGE equations, the
unification of the gauge couplings and some additional assumptions. We use
heterotic boundary conditions (\ref{RGEstring}) for an illustrative purpose,
but the analysis is similar if one uses Type I conditions, (\ref{typeI}).
Below the scale $\mu _{0}$ only the MSSM spectrum and ``running'' apply. The
model is built by assuming that above the scale $\mu _{0}$ only Kaluza Klein
towers for the MSSM gauge bosons exist, and they come in as ${\cal N}=2$
vector multiplets. The ``zero-level'' modes of these Kaluza Klein states are
identified with the corresponding MSSM bosons. To ensure that ``zero-level''
modes are indeed only ${\cal N}=1$ supersymmetric multiplets, additional
symmetry conditions must be imposed and these depend on the type of
manifold. This example corresponds to an orbifold compactification in which
the ${\cal N}=1$ chiral adjoint component of the ${\cal N}=2$ vector
multiplet is odd under the discrete group of the orbifold so that it does
not have zero modes\footnote{%
This ensures that the chiral adjoint field is not present in the low energy
spectrum.}. Furthermore the three generations of the MSSM are assumed to lie
all at the (same) fixed points of the orbifold considered (to avoid the
presence of KK states for these states). In this simple model the
coefficient ${\overline{b}_{i}}$ of (\ref{RGEstring}) is given by 
\begin{equation}
{\overline{b}}_{i}=T_{i}(G)-3T_{i}(G)=\left\{ 0;-4;-6\right\}
_{i}\;\;\;\;\;\;\;i=\{1,2,3\}  \label{bixi}
\end{equation}
where the first and second term account for the chiral adjoint and massless
vector component of the ${\cal N}=2$ vector multiplet, respectively.

Above the scale $\mu _{0}$ the values of $\Delta \gamma _{\phi }$ due to the
excited KK gauge effects are fixed by the following value for $f_{\phi
_{j}}(Q,\mu _{0};\delta )$ \cite{dienes_v1,mondragon} 
\begin{equation}
f_{\phi _{j}}(Q,\mu _{0};\delta )=N(Q,\mu _{0};\delta )-1\;\;\;\;\;\;\;Q>\mu
_{0}
\end{equation}
and a similar expression exists for $g_{H_{u,d}}(Q,\mu _{0};\delta )$. This
follows from the fact that above the scale $\mu _{0}$ only {\it excited} $%
{\cal N}=1$ massless vector KK states contribute (we ignore possible string
form factor effects, eq.(\ref{formfactor})). Using (\ref{RGEstring}) the RGE
equations of the model take the following form (for $\delta=2$, 
\footnote{See
discussion following eq.(\ref{di}) of Section \ref{wchstrings} which
motivates this choice for $\delta$.})
\begin{eqnarray}
\alpha _{i}^{-1}(\mu _{0}) &=&\alpha _{s}^{-1}-\frac{\overline{b}_{i}}{4\pi }%
\ln \left\{ 4\pi ^{2}\,{|}\eta (i){|}^{4}\,\left( \frac{M_{s}}{\mu _{0}}%
\right) ^{2}\,{\bigg\vert}\eta \left[ \frac{i\,3\sqrt{3}\pi }{2e^{1-\gamma }}%
\,\left( \frac{M_{s}}{\mu _{0}}\right) ^{2}\right] {\bigg\vert}^{4}\right\} +%
\frac{b_{i}}{2\pi }\ln \frac{M_{s}}{\mu _{0}}  \nonumber \\
&+&\frac{T_{i}(G)}{2\pi }\ln \frac{\alpha _{s}}{\alpha _{i}(\mu _{0})}+\frac{%
1}{4\pi }\sum_{k=1}^{3}(b_{ik}-2b_{k}\;T_{k}(G)\delta _{ik})\int_{\mu
_{0}}^{M_{s}}\frac{d\ln \mu }{2\pi }\;N(\mu ,\mu _{0},2)\;\alpha
_{k}(\mu )  \label{RGERGE}
\end{eqnarray}
with ${\overline{b}}_{i}$ defined in eq.(\ref{bixi}). If we set $\mu
_{0}=M_{s}$ only the MSSM spectrum and ``running'' would apply below the
unification scale. However, there would be a difference from the familiar
MSSM case because of the string theory boundary condition (eq.(\ref
{RGEstring}) with $\mu _{0}=M_{s}$), with effects on $\alpha _{3}(M_{z})$ to
be discussed later. From the scale $\mu _{0}$ down to the scale $M_{z}$ only
the usual MSSM spectrum and RGE running apply (see \cite{extradim} for
explicit form of the RGE below $\mu _{0}$). We use the values of $\alpha
_{1}(M_{z})$ and $\alpha _{2}(M_{z})$ as a numerical input and keep $%
M_{s}/\mu _{0}$ as a parameter of the model. Using (\ref{RGERGE}) and the
RGE equations below the scale $\mu _{0}$ we can compute the values of $M_{s}$%
, $\alpha _{3}(M_{z})$ and $\alpha _{s}$. The low-energy supersymmetric
thresholds are taken into account as an effective threshold of\footnote{%
A larger, less favoured value for this threshold is $1000$ GeV which further
reduces $\alpha _{3}(M_{z})$.} $300$ GeV which in the MSSM case gives%
\footnote{%
The MSSM value for the strong coupling is computed using two loop running 
{\bf without} any string boundary condition of the type given by eq.(\ref
{RGEstring}) with $\mu _{0}=M_{s}$.} a value for the strong coupling of $%
\alpha _{3}(M_{z})\approx 0.125$, above the experimental limit \cite{wa}, $%
\alpha _{3}(M_{z})=0.119\pm 0.002$.

\begin{table}[tb]
\begin{center}
\begin{tabular}{|c|c|c|l|c|}
\hline
$M_s/\mu_0$ & $M_s$ (GeV) & $\mu_0$ (GeV) & $\alpha_3(M_Z)$ & $\alpha_s$ \\ 
\hline\hline
1.00 & $10^{16}$ & $10^{16}$ & 0.1224 & 0.0411 \\ \hline
1.11 & $6.82\times 10^{15}$ & $6.17\times 10^{15}$ & 0.1212 & 0.0405 \\ 
\hline
1.22 & $4.22\times 10^{15}$ & $3.45\times 10^{15}$ & 0.1197 & 0.0396 \\ 
\hline
1.35 & $2.31\times 10^{15}$ & $1.71\times 10^{15}$ & 0.1180 & 0.0387 \\ 
\hline
1.50 & $1.08\times 10^{15}$ & $7.27\times 10^{14}$ & 0.1158 & 0.0375 \\ 
\hline
1.65 & $4.25\times 10^{14}$ & $2.58\times 10^{14}$ & 0.1133 & 0.0361 \\ 
\hline
1.82 & $1.33\times 10^{14}$ & $7.32\times 10^{13}$ & 0.1103 & 0.0346 \\ 
\hline
2.01 & $3.19\times 10^{13}$ & $1.58\times 10^{13}$ & 0.1068 & 0.0328 \\ 
\hline
2.23 & $5.47\times 10^{12}$ & $2.46\times 10^{12}$ & 0.1028 & 0.0309 \\ 
\hline
\end{tabular}
\end{center}
\caption{{\protect\small {The values of the unification scale $M_s$,
decoupling scale $\protect\mu_0$, the strong coupling at the electroweak
scale and the bare coupling $\protect\alpha_s$ in terms of the ratio $M_s/%
\protect\mu_0$ which is related to the number of additional Kaluza Klein
states approximated by eq.(\ref{number}). Our two-loop results are based on (%
\ref{RGERGE}) and correspond to the case when gauge bosons have Kaluza-Klein
towers. The model fails to increase the unification scale closer to the
heterotic scale, eq.(\ref{mstring}). However, using string boundary
conditions for the RGE and a compactification scale equal to the string
scale (first line in the table) allows one to reduce the strong coupling
from the MSSM value (0.125) to 0.122 which is closer to the experimental
value. This is achieved at the expense of a very small decrease (by a factor
of 2) of the unification scale from the MSSM scale (of $\approx 
2\times 10^{16}$ GeV).}}}
\label{table1}
\end{table}

The numerical results are presented in Table~1. We examined the importance
of the two loop effects above $\mu_0$ on $\alpha_3(M_z)$, relative to the
case when they are neglected and found they are about $3-4\%$. The values of
the unification scale $M_s$ and of the decoupling scale $\mu_0$ are more
sensitive to these effects, their relative variation being $\approx 10\%$.
As the results of Table~1 show, the model fails to give a string scale close
to the weakly coupled heterotic string prediction eq.(\ref{mstring}). 
There is however one important
observation regarding the strong coupling at $M_z$. Assuming the value of $%
\mu_0$ to be equal to the string scale, one finds a value for the strong
coupling $\approx 0.122$, closer to the experimental value than that
predicted\footnote{%
with identical {\it low energy} Supersymmetric (effective) threshold of 300
GeV} by the two loop MSSM of $\approx 0.125$. This effect is due entirely to
the string boundary condition (eq.(\ref{RGEstring})) with $\mu_0=M_s$ and
cannot be computed on pure field theory grounds. The boundary conditions
used in this model considered Kaluza Klein towers for the gauge bosons only
but one could investigate other possibilities which may increase the
unification scale, to bring it closer to the string prediction, eq.(\ref
{mstring}). One can also include Kaluza-Klein towers for the fermions as well
(embedded in ${\cal N}=2$ hypermultiplets) which will enhance the two-loop
contribution above the scale $\mu _{0}$ due to their positive contribution
to ${\overline b}_i$ (which increases the coupling and the higher order
effects). Other constructions are also possible \cite{pq}, where
supersymmetry is broken by a Scherk-Schwarz mechanism to reproduce the
standard model spectrum below the compactification scale\footnote{%
In this section we kept MSSM as the low energy model since we investigated
the case where the compactification scale is very high (see Table~1), 
as the boundary conditions (\ref{RGEstring}) are those of the weakly
coupled heterotic string.}.

\section{Models for low scale unification}\label{type1}

In this section we discuss some aspects of models with KKW states. As it is
well known, in the MSSM one fixes two gauge couplings $\alpha _{1}(M_{z})$
and $\alpha _{2}(M_{z})$ from the experimental input, and with fixed values
for the low energy supersymmetric thresholds one {\it predicts} from the
unification condition the value of $\alpha _{3}(M_{z})$. This only depends
on the symmetry of the model and on the spectrum which fixes the values of
the one loop coefficients, $b_{i}$. One finds at one-loop, up to
Supersymmetric threshold effects 
\begin{equation}
\alpha _{3}^{o-1}(M_{z})=-\frac{b_{2}-b_{3}}{b_{1}-b_{2}}\,\alpha
_{1}^{-1}(M_{z})-\frac{b_{3}-b_{1}}{b_{1}-b_{2}}\,\alpha _{2}^{-1}(M_{z})
\label{MSSMalpha3}
\end{equation}
How does this equation change in models with KKW states? This depends on the 
$({\cal N}=2)$ multiplet content which causes the couplings to run.

\subsection{Power-law running}

We assume the asymmetric case in eq.(\ref{typeI}) leading to the {\it linear}
power-law running in Type~I theories \cite{dudash}. Ignoring a small 
term $\ln(M_s/\mu_0)$), at the one-loop level the 
equations have the form
\begin{equation}  \label{plaw}
\alpha_i^{-1}(M_z)\approx \alpha_s^{-1}+\frac{b_i}{2\pi}\ln\frac{M_s}{M_z} +%
\frac{{\overline b}_i}{12}\frac{M_s}{\mu_0}
\end{equation}
which follows from eqs.(\ref{gauge}), (\ref{typeI}) 
and where ${\overline b}_i$ is the ${\cal N}%
=2$ one loop $\beta$ function coefficient. In this we assume the MSSM ${\cal %
N}=1$ light spectrum (but see below for a discussion of this point).

The relative evolution of the gauge couplings is what determines whether the
couplings unify. We have 
\begin{equation}
\alpha _{i}^{-1}(M_{z})-\alpha _{j}^{-1}(M_{z})\approx \frac{b_{i}-b_{j}}{%
2\pi }\ln \frac{M_{s}}{M_{z}}+\frac{{\overline{b}}_{i}-{\overline{b}}_{j}}{12%
}\frac{M_{s}}{\mu _{o}}
\end{equation}
Clearly if 
\begin{equation}
({\overline{b}}_{i}-{\overline{b}}_{j})=K(b_{i}-b_{j})  \label{lowscale}
\end{equation}
where $K$ is a positive constant, then the MSSM unification will persist
with the replacement $\ln (M_{G}/M_{z})\rightarrow \ln (M_{s}/M_{z})+2\pi
K(M_{s}/\mu _{0})/12$. The fact that the MSSM is consistent with unification
with $M_{G}\approx 3\times 10^{16}$ GeV gives a value of unification scale
for our example of $M_{s}/\mu _{0}\approx 63/K$. One can also determine the
string coupling in terms of the value $\alpha _{g}$ of the unified coupling
in the MSSM ($\alpha _{g}\approx 1/24$) to find $\alpha _{sg}^{-1}=\alpha
_{g}^{-1}-\sigma M_{s}/(12\,\mu _{0})$ where $\sigma ={\overline{b}}%
_{i}-Kb_{i}$.

Given $\alpha _{sg}$ and $M_{s}/\mu _{0}$ we may determine whether higher
order radiative corrections are under control. Following the discussion of
Section \ref{pe} we see that the corrections to \ the running coupling at $%
O(\alpha ^{2})$ are calculable provided $\alpha _{s}N_{KK}/(4\pi )<1.$ Using 
$N_{KK}\approx 2M_{s}/\mu _{0}\approx 126/K$ this requires $10\alpha
_{sg}/K<1.$ Whether or not this is true is model dependent but it will be
the case for a small unified coupling $\alpha _{sg}\approx \alpha _{g}.$
The reason higher order corrections may remain under control is that there
are a relatively small number of Kaluza Klein modes below the string scale.

Of course the important question is whether condition (\ref{lowscale}) can
be satisfied ? In the MSSM the one loop coefficient $b_{i}$ has the
following form $b_{i}=\sum_{\psi }^{{}}T_{i}(\psi )-3T_{i}(G)$ where the sum
runs over all matter fields of the MSSM and the second term accounts for the
gauge sector. If all the MSSM states have KK towers, the coefficients ${%
\overline{b}}_{i}$ (i=1,2,3) have such values that condition (\ref{lowscale}%
) is not satisfied. Indeed, in this case ${\overline{b}}_{i}$ is given by ${%
\overline{b}}_{i}=2\sum_{\psi }^{{}}T_{i}(\psi )-2T_{i}(G)$ where the sum is
over the MSSM matter fields which have all KK towers and the factor 2 in
front accounts for their embedding in ${\cal N}=2$ hypermultiplets.
Similarly, the second term in the expression of ${\overline{b}}_{i}$
accounts for the gauge sector with KK states being ${\cal N}=2$ vector
supermultiplets (which contain a ${\cal N}=1$ massless vector and a chiral
adjoint multiplet), hence the factor 2 which multiplies it. Thus ${\overline{%
b}}_{i}=2b_{i}+4T_{i}(G)$ and (\ref{lowscale}) is not satisfied.

One way out of this difficulty is to include further KK states which
together with the KK excitations of the MSSM spectrum will have one loop
coefficients which satisfy (\ref{lowscale}). An example of this class was
found by Kakushadze \cite{zurab1} who constructed a simple $Z_2$ orbifold
model which breaks ${\cal N}=2$ to ${\cal N}=1$. The original ${\cal N}=2$
spectrum is that of MSSM KK excitations together with those of additional
superfields $F_{\pm}$. These are $SU(3)\times SU(2)$ singlets with $U(1)_Y$
charge $\pm 2$ respectively. For this model condition (\ref{lowscale}) is
indeed satisfied with $K=1$. However the model is not quite what is desired
because it contains additional light ${\cal N}=1$ fields with the quantum
numbers of $F_{\pm}$ and as a result we find 
\begin{equation}  \label{Z1}
\alpha_3^{-1}(M_z)=-\frac{b_2-b_3}{b_1-b_2}\alpha_1^{-1}(M_z) -\frac{b_3-b_1%
}{b_1-b_2}\alpha_2^{-1}(M_z) +\frac{1}{2\pi}\frac{6}{7}\ln\frac{M_s}{M_F}
\end{equation}
where the last term is the one-loop contribution of $F_{\pm}$ states to the
running of the gauge couplings. This term depends on the (unknown) ratio $%
M_s/M_F$ and thus eq.(\ref{Z1}) makes no prediction for $\alpha_3(M_z)$
following from the unification condition. A detailed mechanism or further
input is needed to fix the value $M_s/M_F$. Thus, although the model may
allow the presence of a low (TeV) string scale, it does not {\it predict} $%
\alpha_3(M_z)$.

For the purpose of exploring the most realistic possibility for unification
with a low string scale we will assume the {\it absence} of the ${\cal N}=1$
states $F_{\pm }$ (this could perhaps be arranged by a different orbifold
construction). With this minimal spectrum we can achieve power-law
unification following from eq.(\ref{plaw}). However there is a potential
problem due to the sensitivity of such power-law unification to the
thresholds. To see this, note that in eq.(\ref{plaw}) we have assumed the
same thresholds ${\cal T}\equiv M_{s}/\mu _{0}$ for all three gauge group
factors. ${\cal T}$ is determined by the high scale cut-off $M_{s}$ and the
low scale cut-off $\mu _{0}$. As discussed in Section \ref{eft} the latter
is the ``bare'' mass and is not sensitive to radiative corrections in the
supersymmetric limit. However it is sensitive to contributions from SUSY
breaking and other spontaneous symmetry breaking sectors which are likely to
be flavour dependent effects giving non-universal ${\cal T}$. Further, as
discussed above, the high scale cut-off $M_{s}$ is determined by the
cross-channel exchanges. We have no reason to think that the coupling of the
direct channel open string states to the closed channel exchange processes
is not sensitive to higher order radiative corrections and SUSY breaking.
Such effects are likely to give a non-universal threshold ${\cal T}$ in eq.(%
\ref{plaw}). Given this it is important to determine how sensitive the
predictions are to the assumption of universal thresholds in eq.(\ref{plaw}%
). Since $\alpha _{3}(M_{z})$ is the most poorly determined of the three
gauge couplings, we concentrate on it. Requiring unification should
reproduce the observed value of $\alpha _{3}(M_{z})$ to an accuracy of $%
\delta \alpha _{3}$, we find 
\begin{equation}
\frac{\delta {\cal T}}{{\cal T}}\approx \frac{12}{{\overline{b}}_{3}{\cal T}}%
\delta \alpha _{3}^{-1}\approx \frac{1}{15}\delta \alpha _{3}^{-1}
\end{equation}
Given that\footnote{%
We take the accuracy $\delta \alpha _{3}(M_{z})\approx 0.006$ roughly equal
to the two loop contribution in the MSSM.} $\delta \alpha _{3}^{-1}\approx
0.5$ we see that ${\delta {\cal T}}/{\cal T}\approx {1}/{30}$. This limit
suggests we should choose the scales at least a factor of 30 above the SUSY
breaking scale to avoid having an unacceptable sensitivity to SUSY breaking.
This corresponds to $\mu _{0}>10^{4}-10^{5}$ GeV and an associated string
scale $M_{s}>10^{6}-10^{7}$ GeV. Thus we see that power-law running avoids
the threshold fine tuning problem only if the string scale is relatively
large.

\subsection{Logarithmic running}\label{logrun}

An interesting possibility \cite{dudash,bachas} for unification in
Type I$^{\prime }$ theories occurs if the compactification 
radii are equal. Then
from eqs.(\ref{gauge}),(\ref{typeI}) we find  
\begin{equation}
\alpha _{i}^{-1}(M_{z})=\alpha _{s}^{-1}+\frac{b_{i}}{2\pi }\ln \frac{M_{s}}{%
M_{z}}+\frac{\overline{b}_{i}}{2\pi }\ln \frac{\mu _{0}}{M_{s}}
\label{logrunning}
\end{equation}
where ${\overline{b}_{i}}$ is the ${\cal N}=2$ one-loop beta function
coefficient. The presence of the (mild) logarithmic running would suggest
that  unification at a high scale ($\mu_0$) above $M_s$ is possible
while keeping the string scale low (TeV region).
Here we investigate this possibility.
If eq.(\ref{lowscale}) is true then the usual MSSM unification
will occur but with $(\mu _{0}/M_{s})^{K}M_{s}\approx 10^{16}$ GeV. For the
case $K=1$ the unification occurs at $10^{16}$ GeV.

What is the threshold sensitivity in this case? From eq.(\ref{logrunning})
we find $\delta {\cal T}/{\cal T}\approx 2\pi \delta \alpha _{3}^{-1}/{%
\overline{b}_{3}}\lower3pt\hbox{$\, \buildrel < \over \sim \, $}{\cal
O}(1)$. However this apparent relaxation of the fine tuning requirement is
misleading because here the string cut-off scale $\mu _{0}\approx
10^{16} \,GeV$ corresponds to a cross channel exchange state with
mass $M_{s}^{2}/\mu _{0}$. For a 1 TeV string scale this is $10^{-10}$ GeV
and corresponds to the mass of the KK modes in the closed string channel.
However the exchanged state controlling the UV behaviour of the open string
channel has vacuum quantum numbers. Such a state has no symmetry (local
gauge symmetry etc) to protect it from receiving contributions to its mass
at scales below the composite scale. As we have seen the KK modes have
pointlike couplings up to the string scale and so we expect that such a
state with vacuum quantum numbers will acquire mass from supersymmetry
breaking and other symmetry breaking sectors. Thus we consider it more
likely that for the case under discussion the UV divergence in the open
string channel will be controlled by an exchange particle with a mass close
to the supersymmetry breaking scale i.e $1\,TeV\approx M_{s}^{2}/\mu _{0}$. If
this is the case logarithmic running will not be a viable mechanism for
gauge unification (with $M_s\approx 1\, TeV$) 
because $(\mu _{0}/M_{s})\approx 1$ and so there is no 
large log in eq.(\ref{logrunning}).

\section{Conclusions}

In this paper we have considered whether string threshold corrections can be
approximately described by a effective field theory valid below an UV cutoff
determined by the underlying string structure. In the weakly coupled
heterotic string case the UV cutoff is at the string scale. Below this scale
there are only a finite number of degrees of freedom and so one can
approximate the string calculation using an effective field theory in which
one included all modes with mass less than the string scale and assuming
pointlike coupling for these modes. In the case of type I/I$^{\prime }$
theories the UV cutoff is much larger than the string scale for the
contributions of Kaluza Klein (winding ) modes. However in this case too the
effective field theory approach may be used because, due to supersymmetry,
the string states do not contribute to the radiative corrections of gauge
couplings. As a result only a finite number of massless states (before
``Higgsing'') contribute together with a finite number of Kaluza Klein
(winding) modes. Just as for the heterotic string case, except for the
states close to the cut-off scale, these states have pointlike coupling and
so an effective field theory approach is again possible.

We found that the most significant error to the calculation of gauge
coupling evolution in the effective field theory approach came from the
assumption of pointlike couplings for the Kaluza Klein states close to the
string scale. However the effect of this approximation can be absorbed in a
change of the cut-off scale. As a result the interpretation of the cut-off
scale as the string scale has a significant error but the other features of
the string threshold effects on gauge couplings are well described by the
effective field theory approach. Moreover the effective field theory offers
a convenient way to determine the structure of higher order radiative
corrections in these theories. If detailed information is needed about the
string scale the effective field theory calculation can be modified to take
account of the non-pointlike couplings of the states close to the string
scale and we presented a simple way to do this. 

The string threshold effects show that the gauge couplings can have power
law running in the case of large compactification scale although the
effective number of ``decompactified'' dimensions is model dependent. This
raises the interesting possibility of achieving unification of gauge
couplings at a low scale, something that is desirable in theories in which
the string scale is itself small. However the power law running occurs only
in the ${\cal N}=2$ sector and hence is governed by the gauge quantum
numbers in this sector. As a result in such models the successful prediction
of the usual MSSM running for gauge couplings (which comes from the ${\cal N}%
=1$ sector) must be viewed as an accident. Moreover, while it is possible to
choose the ${\cal N}=2$ sector to get a low scale of unification via power
law running, in order to avoid a fine tuning problem associated with the
threshold of the ${\cal N}=2$ states, it is necessary to have a relatively
high string and compactification scale, $>{\cal O}(10^{6}\,GeV)$. We also
examined the possibility of a theory with a low string scale but with a high
scale of unification through logarithmic running. However this scheme seems
to suffer from a severe fine tuning problem associated with the need for
extremely light closed string states carrying vacuum quantum numbers. In the
absence of a symmetry protecting these states we expect them to be driven to
the supersymmetry breaking scale spoiling the mechanism needed to have
logarithmic running to a very high scale.  Finally, while unification
through power law running is possible in theories with a low
compactification and string scale, it seems  much more contrived than the
original proposal of logarithmic running and unification at a high scale.

\section{Acknowledgments}

The authors thank A. Lukas and F. Quevedo for useful discussions. D.G.
acknowledges the financial support from the part of the University of Oxford
(Leverhulme Trust research grant). The research is supported in part by the
EEC under TMR contract ERBFMRX-CT96-0090.

\end{document}